\documentclass[11pt,a4paper]{article}
\usepackage{jinstpub}
\usepackage{natbib}
\usepackage{lineno}
\usepackage{upgreek} 
\newcommand{\rn}{$^{222}$Rn}
\newcommand{\ra}{$^{226}$Ra}
\newcommand{\bi}{$^{214}$Bi}
\newcommand{\po}{$^{214}$Po}
\newcommand{\refftpc}{$R^{\text{eff}}_{\text{TPC}}$} 
\newcommand{\reff}{$R^{\text{eff}}$} 

\graphicspath{{Figures/}}

\title{Observation of Radon Mitigation in MicroBooNE by a Liquid Argon Filtration System}
\collaboration{MicroBooNE Collaboration}
\date{August 2021}
\begin{document}

\author[hh]{P.~Abratenko}
\author[d]{J.~Anthony}
\author[s]{L.~Arellano}
\author[gg]{J.~Asaadi}
\author[ee]{A.~Ashkenazi}
\author[k]{S.~Balasubramanian}
\author[k]{B.~Baller}
\author[u]{C.~Barnes}
\author[x]{G.~Barr}
\author[t,ee]{J.~Barrow}
% ...

% Institutions in alphabetical order
% ...
\affiliation[d]{University of Cambridge, Cambridge CB3 0HE, United Kingdom}
% ...
\affiliation[k]{Fermi National Accelerator Laboratory (FNAL), Batavia, IL 60510, USA}
% ...
\affiliation[s]{The University of Manchester, Manchester M13 9PL, United Kingdom}
\affiliation[t]{Massachusetts Institute of Technology (MIT), Cambridge, MA, 02139, USA}
\affiliation[u]{University of Michigan, Ann Arbor, MI, 48109, USA}
% ...
\affiliation[x]{University of Oxford, Oxford OX1 3RH, United Kingdom}
% ...
\affiliation[ee]{Tel Aviv University, Tel Aviv, Israel, 69978}
% ...
\affiliation[gg]{University of Texas, Arlington, TX, 76019, USA}
\affiliation[hh]{Tufts University, Medford, MA, 02155, USA}
% ...

\emailAdd{microboone\_info@fnal.gov}

\date{\today}
\abstract{
The MicroBooNE liquid argon time projection chamber (LArTPC) maintains a high level of liquid argon purity through the use of a filtration system that removes electronegative contaminants in continuously-circulated liquid, recondensed boil off, and externally supplied argon gas. We use the MicroBooNE LArTPC to reconstruct MeV-scale radiological decays. Using this technique we measure the liquid argon filtration system’s efficacy at removing radon. This is studied by placing a 500~kBq \rn~source upstream of the filters and searching for a time-dependent increase in the number of radiological decays in the LArTPC. In the context of two models for radon mitigation via a liquid argon filtration system, a slowing mechanism and a trapping mechanism, MicroBooNE data supports a radon reduction factor of greater than 99.999\% or 97\%, respectively. Furthermore, a radiological survey of the filters found that the copper-based filter material was the primary medium that removed the \rn. This is the first observation of radon mitigation in liquid argon with a large-scale copper-based filter and could offer a radon mitigation solution for future large LArTPCs.
}
\maketitle

\section{Introduction}

Detectors that employ noble liquids as an active medium have come to the forefront of particle physics at the MeV energy scale~\cite{exoresults,DarkSide,gerdaFinal,legend,argoMeV,benemev,deap,xenon,pandax}, providing low thresholds and precise measurements of the positions and energies of interacting particles. One challenge when deploying liquid noble detectors to explore physics at MeV and lower energies is the necessity to maintain an environment low in radioactivity, as decays can mimic the signals being studied. One major background is the decay of radon-222 and its decay products. In deep underground detectors, this contamination has been kept to a level on the order of 2~$\upmu$Bq/kg~\cite{dsrnrates} to maintain backgrounds to an acceptable rate.

MicroBooNE is an 85~metric ton liquid argon time projection chamber (LArTPC) that operated from 2015 to 2021~\cite{uBdet}. LArTPCs operate by collecting ionization electrons and scintillation light produced by charged particles traversing the liquid argon medium. The measurement presented in this paper leverages only this charge signal. These ionization electrons drift in an applied electric field toward a charge-sensitive anode plane creating signals that are sampled and digitally recorded. In MicroBooNE, the applied drift field is 273 V/cm~\cite{E_cosmics,E_laser} and the maximum drift distance is 2.56~m. This corresponds to an average drift velocity of 1.098~mm/$\upmu$s \cite{E_laser} for electrons and 4~mm/s for argon
ions \cite{E_cosmics}.
MicroBooNE's charge readout is achieved using an array of readout wires arranged into three planes and connected to waveform digitization electronics. This analysis uses the collection plane which has an effective noise charge (ENC) of 300~electrons~\cite{ubnoise}.
The ability for LArTPCs to detect this charge depends on the ability to drift ionization over long distances through liquid argon (LAr).  While drifting, the ionization electrons can be absorbed by electronegative contaminants in the LAr, which attenuates the measured signal at the anode. This attenuation appears as an exponential decrease in signal size as a function of the drift time, with the characteristic time constant commonly referred to as the ``electron lifetime.'' The electron lifetime measured in MicroBooNE during normal operations is in excess of 20~ms~\cite{ubcalib}, such that any signal attenuation is negligible over the maximum drift time of 2.3~ms.

Generally, liquid noble experiments searching for dark matter~\cite{DarkSide20k} and rare processes~\cite{lux,xe1t,xentrn} employ filtration in the gaseous phase, passing the recirculated gases through activated charcoal and other filter materials to remove contaminants that can attenuate signals or produce radioactive background signals. Instead, MicroBooNE filters in the liquid phase to remove electronegative contaminants employing a two-part filtration system, known as a ``filter skid''~\cite{uBdet}. First, gaseous argon which has boiled off the LAr in the cryostat is re-condensed and mixed with recirculated LAr drawn from the cryostat in the liquid phase. This mixture is then pumped at a typical rate of 0.63~L$/$s through a 4\AA~molecular sieve~\cite{molesieve} (mole sieve) filter that removes some contaminant molecules, such as water. The liquid
is then passed through a second filter containing BASF CU-0226 S, which is a high--surface area pelletized alumina impregnated with copper~\cite{copper};
this stage was designed to remove residual oxygen. Together these two filters are highly efficient at removing electronegative impurities from the LAr, allowing MicroBooNE to maintain electron lifetimes significantly above the 3~ms design specification~\cite{ubcalib}. This system filters the entire volume of the MicroBooNE cryostat roughly every three days. The MicroBooNE cryogenic system includes two independent filter skids and it is possible to switch between them, allowing maintenance of either filter without interrupting detector operations~\cite{uBdet}. The first set of filters, referred to as the full-sized filter skid, is the primary system used during MicroBooNE operations and contains 77~L of each filtering material. The second, referred to as the 30\%-sized filter skid, contains 25.2~L of mole sieve material and 23.5~L of copper material.

\begin{figure}
    \centering
    \includegraphics[width=0.8\linewidth]{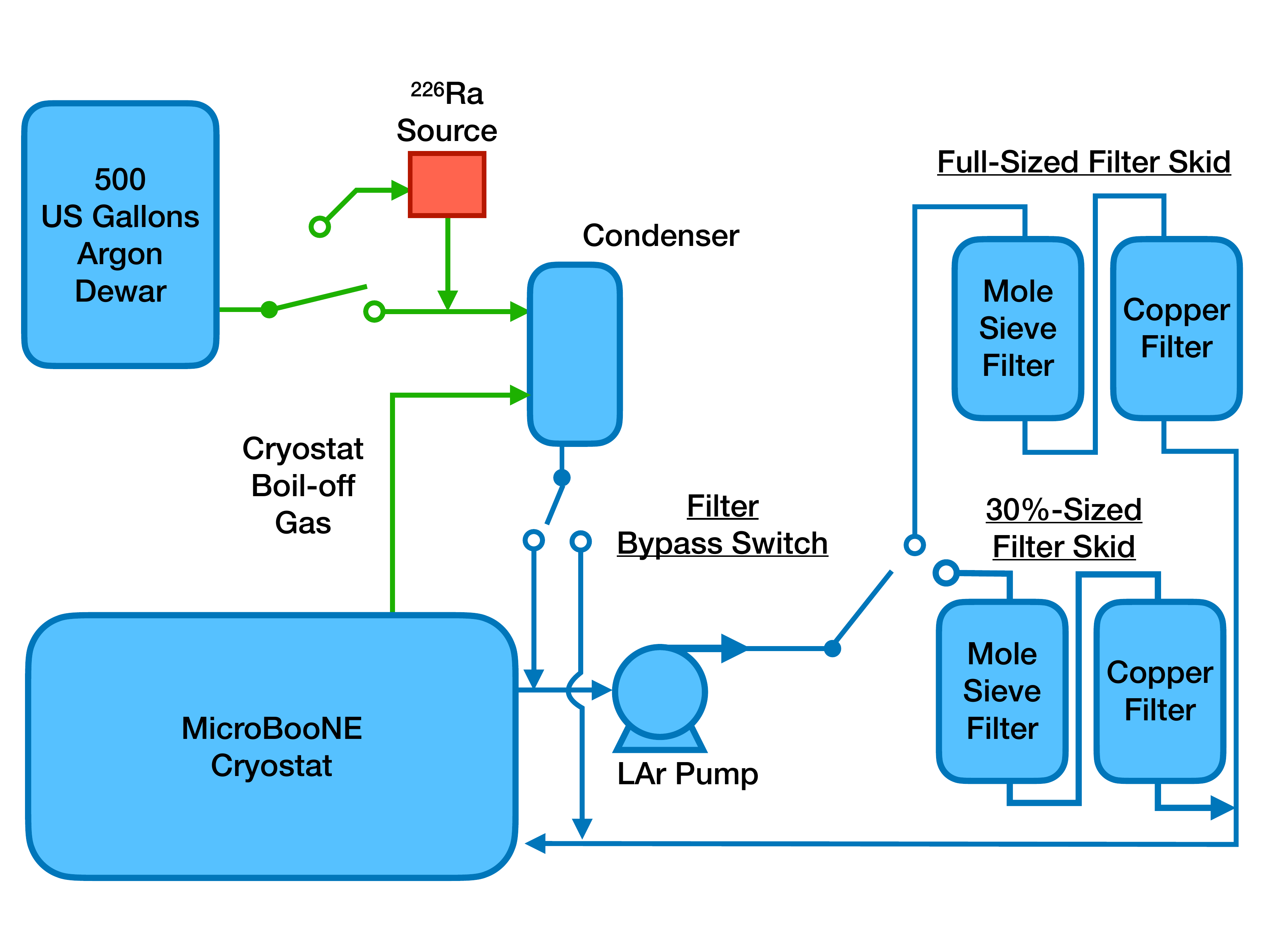}
    \caption{A flow diagram of the MicroBooNE cryogenic system, which was modified for these studies to include a~\ra~source (designated in red above); green lines indicate gas argon flow and blue lines indicate argon liquid flow. The arrows designate the flow of liquid and gaseous argon. The ``filter bypass switch'' enables recondensed argon to flow past the liquid argon pump and filter skids and flow directly back to the cryostat. The switch downstream of the liquid argon pump allows the liquid argon to flow either through the full-sized filter skid or the 30\%-sized filter skid before flowing back into the cryostat. The orientation of the switches is shown as closer toward their default configuration during standard operation.}
    \label{fig:cryo}
\end{figure}

While these filters have demonstrated their ability to remove electronegative contaminants, their effectiveness in mitigating radioactive contaminants has not previously been explored. To study this, a Pylon Electronics, Inc.\ Model 2000A~\cite{pylon} 500~kBq \ra~source was placed inline with gas flow from an external 500~US~gallon argon dewar (used to regularly replenish the cryostat), in the configuration shown in figure~\ref{fig:cryo}. The \ra~($t_{1/2}=1600$~y) contained within this source decays, producing \rn~($t_{1/2}=3.82$~d) gas which mixes with the gaseous argon. This radon-doped gaseous argon is then passed into the condenser and mixed with the recirculated LAr. This LAr then passes first through a mole sieve filter followed by a copper filter before being reintroduced into the MicroBooNE cryostat.  

\section{Radon-222 Decay Topology}
\label{sec:topo}

To identify \rn~decays in the MicroBooNE LArTPC we leverage a unique signature of its decay chain: the \bi-\po~decay series. This sequence of decays occurs in the ${^{238}\mathrm{U}}$ series, and proceeds as follows:
\[
{{}^{214}_{83}\mathrm{Bi}}~
\xrightarrow[19.9~\text{min}]{\beta^-,~Q=3.3~\text{MeV}}~
{{}^{214}_{84}\mathrm{Po}}~
\xrightarrow[164.3~\upmu\text{s}]{\alpha,~7.7~\text{MeV}}~
{{}^{210}_{82}\mathrm{Pb}}
\]
The 164~$\upmu$s half life of \po~equates to a roughly 10~cm separation between the drifting ionization electrons from the \bi~and \po~decay products. Meanwhile, the recoil distance of the heavy \po~ion and its drift during this same period of time are much smaller than the mm-scale position resolution in MicroBooNE. This means that the decay products of \bi~and \po~will be detected on the same channel (wire), but clearly separated in time, creating a powerful topological handle for event selection.

In addition to the characteristic spatial separation of low-energy signals, we can also leverage the particles that result from the radioactive decays. The \bi~decays via $\beta$-decay with the predominant transition having an end-point of 3.269~MeV (producing $\approx100,000$~electrons) while the \po~decays via a monoenergetic 7.686~MeV $\alpha$~emission (producing $\approx4,000$~electrons)~\cite{nds}. The distance traveled by the $\beta$ emerging from the \bi~decay will span between 1 to 4 channels of the MicroBooNE readout, while the $\alpha$ will typically be constrained to a single channel. When $\alpha$ particles deposit their energy in LAr they tend to produce relatively little ionization charge and more scintillation light~\cite{Mei:2007jn}; for the $\alpha$ produced in the signal \po~decays, the scintillation yield is $\sim95$\% larger than the ionization yield~\cite{nestbench}. Thus, topologically we search for a small electron-like energy deposition in the 1--3~MeV range followed by a faint point-like energy deposition on the same channel and separated by several centimeters in the drift direction. In most cases the \bi~decay also produces a coincident $\gamma$, which can result in additional isolated low-energy activity in the vicinity of the decay; this activity is not used in our selection.

\section{Simulation and Selection}\label{sec:sim}

\begin{figure}
    \centering
    \includegraphics[width=0.8\linewidth]{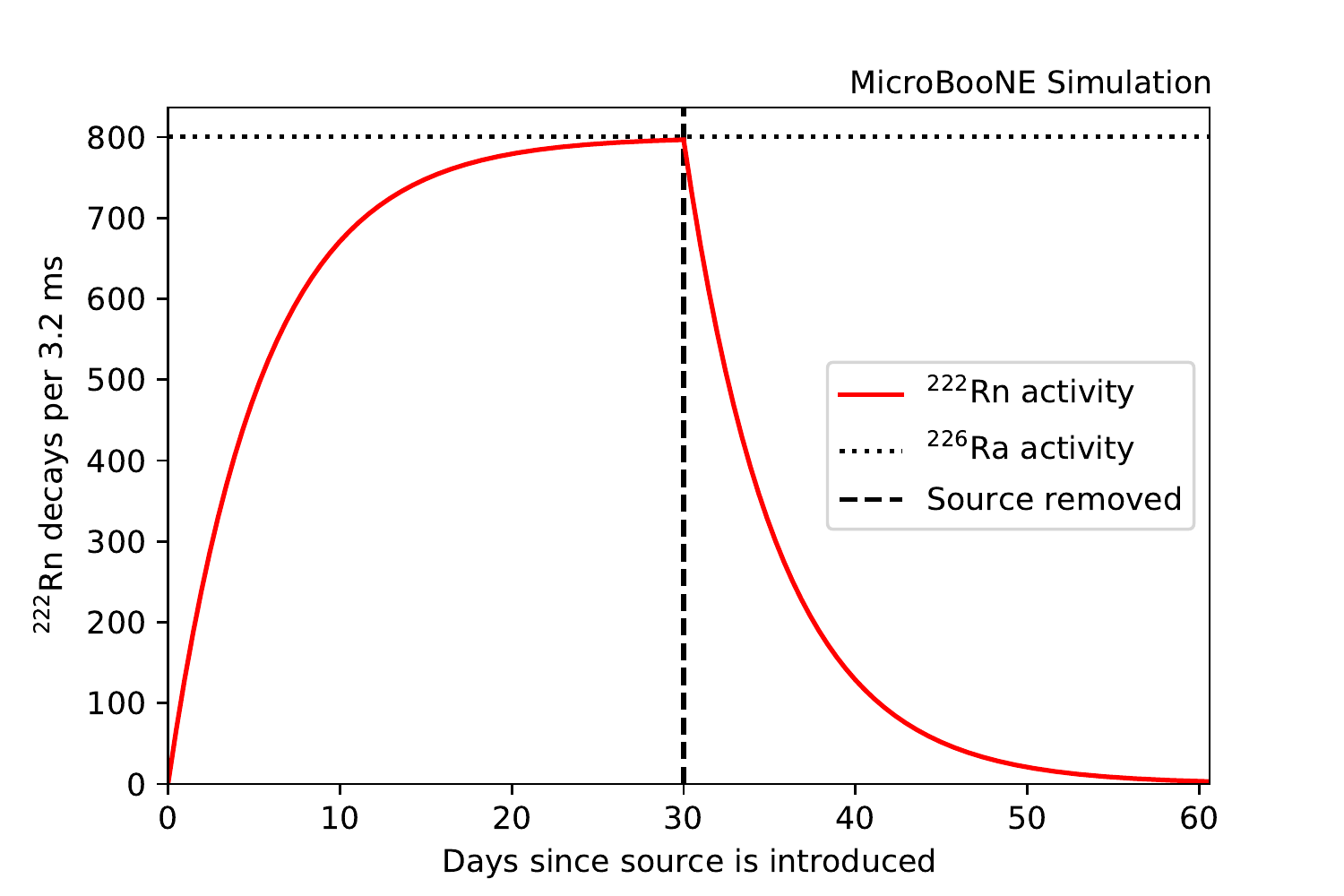}
    \caption{The number of $^{222}$Rn decays expected per 3.2~ms detector readout as a function of time since the source was added under the assumption that the cryostat activity comes into secular equilibrium with that of the source. This also assumes that the filters remove none of the injected $^{222}$Rn. In this example, we then model removing the source at 30~days and study the reduction in the number of \rn~decays.}
    \label{fig:rates}
\end{figure}

To determine the amount of \rn~introduced into the MicroBooNE cryostat we start with two assumptions: the activity in the cryostat will reach secular equilibrium with the source activity; and the filtration system will remove no radon. Then we estimate the number of $^{222}$Rn decays expected in the active volume of the MicroBooNE TPC during a single 3.2~ms readout for our 500~kBq source, shown in figure~\ref{fig:rates}. We find that if the radon is allowed to flow freely into the detector we would expect to see $>\!\!100$ $^{222}$Rn decays per readout within the first day the source is introduced, and after two weeks we would expect to see roughly 700 $^{222}$Rn decays per readout. The functional form of the activity estimate is described in further detail in section~\ref{sec:efficacy}.
 
\begin{figure}
    \centering
    \includegraphics[width=0.8\linewidth]{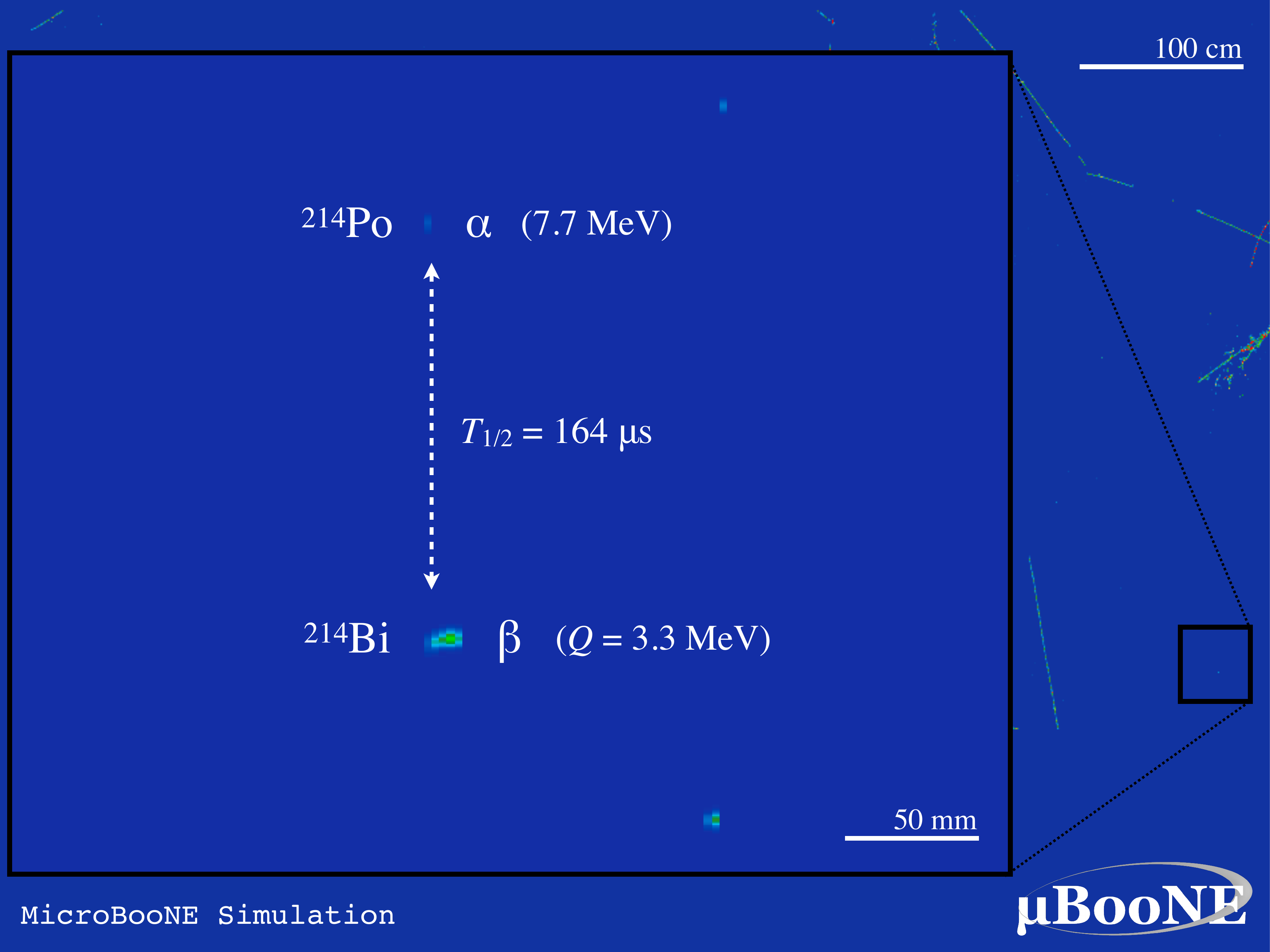}
    \caption{A simulated \bi--\po~decay event, overlaid with MicroBooNE cosmic ray-only data, labeled to indicate the associated $\beta$~and $\alpha$ activity. The apparent separation between the $\beta$-decay and $\alpha$-decay products comes from the ionization charge of the $\beta$ drifting towards the anode-plane over the 164~$\upmu$s \po~half-life. No other activity in the event display meets the signal topology requirements. The scale bars shown apply both horizontally and vertically.}
    \label{fig:bipo-evd}
\end{figure}

Using these estimates we can build a detailed simulation of these decays. We employ the \textsc{Decay0} event generator~\cite{decay0,decay4} to simulate the coincident decay of \bi~and \po. The number of decays is randomized and selected based on the source's activity. These decays are then overlaid with MicroBooNE cosmic data, taking into account the full detector response model and calibrations~\cite{ubcalib,E_cosmics,E_laser}. An example simulated \bi--\po~decay is shown in figure~\ref{fig:bipo-evd}.

Both data and simulated events then follow an identical reconstruction path. We start by processing the raw detector signals with a series of low-threshold signal processing algorithms, which removes the electronics noise and accounts for the field responses near the wire planes~\cite{signal,signal2}. This process results in Gaussian-shaped signals that are subsequently fit with normal distributions creating what are known as ``hits.'' These hits represent an energy deposition on a channel at a point in space in the drift coordinate. For isolated charge deposits exhibiting relatively little wire activity, contiguous hits are collected into clusters~\cite{avinay,argoMeV}.
Since MicroBooNE sits near the Earth's surface, it is constantly inundated by cosmic ray activity~\cite{ubcosmics}. To remove low-energy activity associated with these cosmic rays, we exclude activity occurring within a cylindrical region of radius 15~cm centered around the trajectory of track-like objects~\cite{avinay,argoMeV}.

The remaining clusters, around 270/event on average (or 0.992 events per kg per second in the TPC active volume), are then evaluated to determine if they fit our criteria as \bi--\po~coincident decay candidates, which are based on the topological features described above. Specifically, the selection identifies two separate small energy deposits (with fewer than five hits) which either start or end on the same readout channel. We require that these two energy deposits be within 6.25~$\upmu$s to 312.5~$\upmu$s of each other, and the second cluster must have less ionization charge than the first. 
When applied to simulated events, the efficiency of the event selection -- as defined by correctly tagging the $\beta$-decay -- is found to be 46\%. 
To assess an uncertainty on this efficiency, we take as due to the selection efficiency the full difference between the observed and predicted rates of candidate events early in the filter-bypass configuration (discussed in section~\ref{sec:runconfig}). This data-driven technique accounts for uncertainties in our modeling of particle decays, detector response, source efficiency, and reconstruction uncertainties.  
This procedure yields a final selection efficiency estimate of $(46^{+31}_{-29})\%$.
According to the simulation, the primary source of inefficiency in the selection is the non-identification of the \po~$\alpha$-signals originating far from the wire planes due to longitudinal diffusion~\cite{diff} of ionization electrons reducing the signal below threshold.

\section{Run Configurations and Findings}
\label{sec:runconfig}
\begin{figure}
    \centering
    \includegraphics[width=0.8\linewidth]{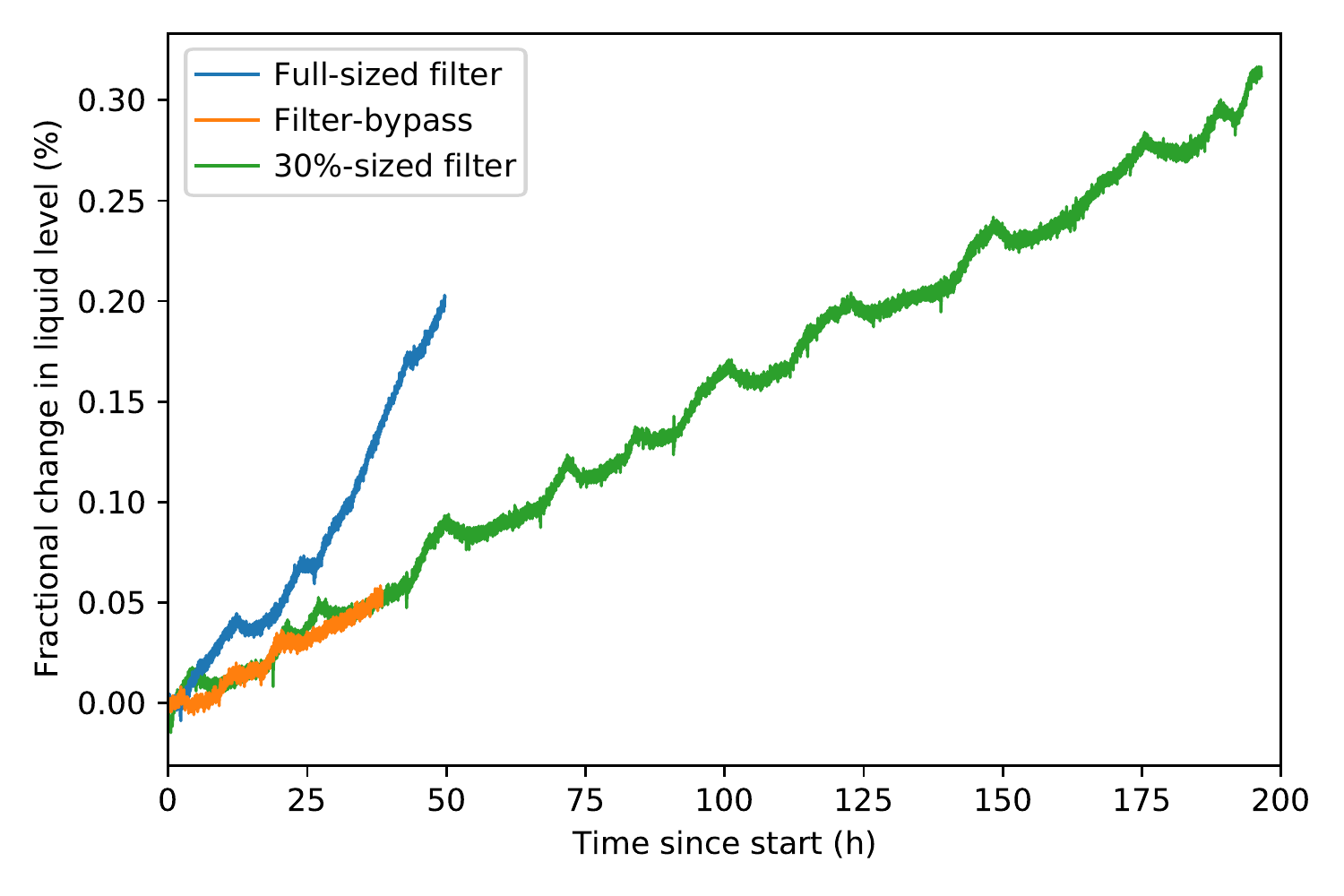}
    \caption{Fractional increase in the cryostat LAr levels during each run configuration.}
    \label{fig:level}
\end{figure}

In this study the MicroBooNE cryogenic system was operated in three configurations:
\begin{enumerate}
    \item Filter-bypass: the recondensed gas was not passed through the filters but the recirculated LAr was.
    \item Full-sized filter: the recondensed gas and recirculated LAr were passed through the 77~L (53.1~kg) mole sieve filter followed by the 77~L (65.5~kg) copper filter before entering the cryostat.
    \item 30\%-sized filter: the recondensed gas and recirculated LAr were passed through the 25.2~L (17.4~kg) mole sieve followed by the 23.5~L (20~kg) of copper before entering the cryostat.
\end{enumerate}
In each of these modes, we constantly flowed gaseous argon from an external 500~US~gallon dewar over our \ra~source and into the cryostat.
The cryostat liquid level was continuously monitored, and the relative change during each period is shown in figure~\ref{fig:level}. The increase in the LAr level does not appreciably change the mass of LAr in the cryostat or impact detector operation, except during the filter bypass when the unfiltered LAr carried electronegative contaminants into the cryostat. This figure highlights that the radon-laced carrier gas was able to make it into the cryostat.

\begin{figure}
    \centering
    \includegraphics[width=0.8\linewidth]{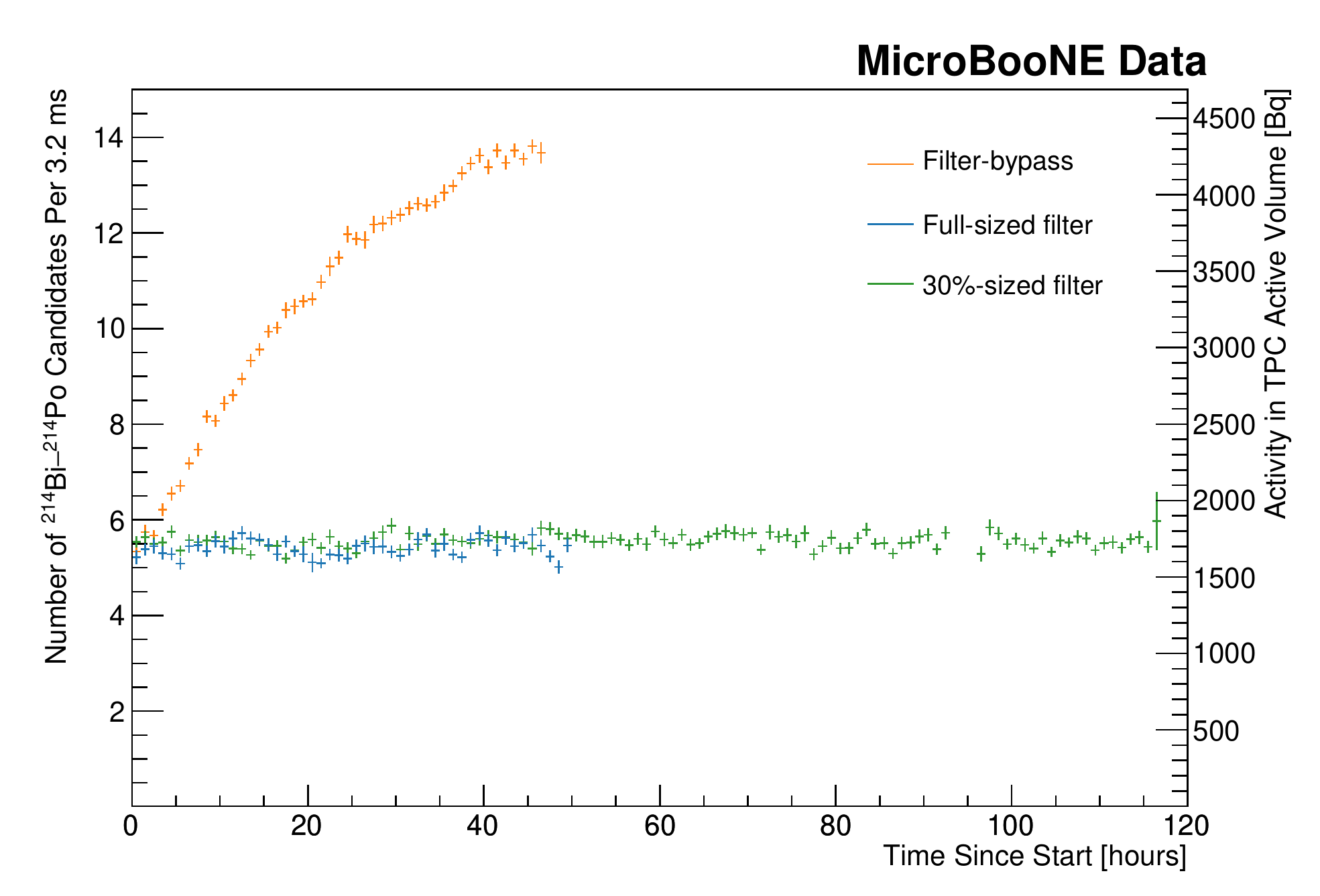}
    \caption{The average number of \bi--\po~coincident decays identified per readout (3.2~ms) versus time since that running condition was entered. This activity rate can also be expressed in terms of Bq in the TPC active volume, shown on the right $y$-axis. The leveling off of the candidates observed during the filter-bypass run is a result of reduced selection efficiency brought on by increasing electronegative contamination.}
    \label{fig:radondata}
\end{figure}

A cross-check of the radon doping system was performed by running in filter-bypass mode. Here, the radon-doped condensed LAr was directed around the filter system and injected directly into the cryostat's liquid phase. Running in this mode allowed us to verify that the radon doping system was capable of bringing radon-doped LAr into the detector volume, and provided a test of the event selection and identification of \bi--\po~candidates. However, bypassing the filtration system also stops the removal of electronegative contaminants, degrading the electron lifetime. The resulting spatial non-uniformities across drift distance complicate the analysis, and in particular, make it difficult to directly establish the source activity since the true position of any given decay in the drift direction is unknown. The filter-bypass data was taken after the full-sized filter data and before the 30\%-sized filter data. The argon electron lifetime was allowed to recover to within 2\% of the nominal value before data-taking with the 30\%-sized filter was started. Figure~\ref{fig:radondata} shows the number of \bi--\po~candidates that were selected as a function of time for all three configurations studied. These distributions will be composed of a flat distribution, that corresponds to the steady-state backgrounds (such as cosmic backgrounds, noise features, or radiological decays) for this analysis, and a rising component from the injected radon source, predicted by the distribution in figure~\ref{fig:rates}. The analysis demonstrates that when employing either of the MicroBooNE filter skids, no appreciable activity from the \ra~chain, up to and including \bi--\po, made its way into the detector. The leveling off of the \bi--\po~candidates observed at 40~h for the filter-bypass data observed in figure~\ref{fig:radondata} is not a result of reaching secular equilibrium, as shown in figure~\ref{fig:rates}, but instead results from the degradation of argon purity due to bypassing the filters.

\section{Radiological Survey of Filters}

\begin{figure}
    \centering
    \includegraphics[width=0.5\linewidth]{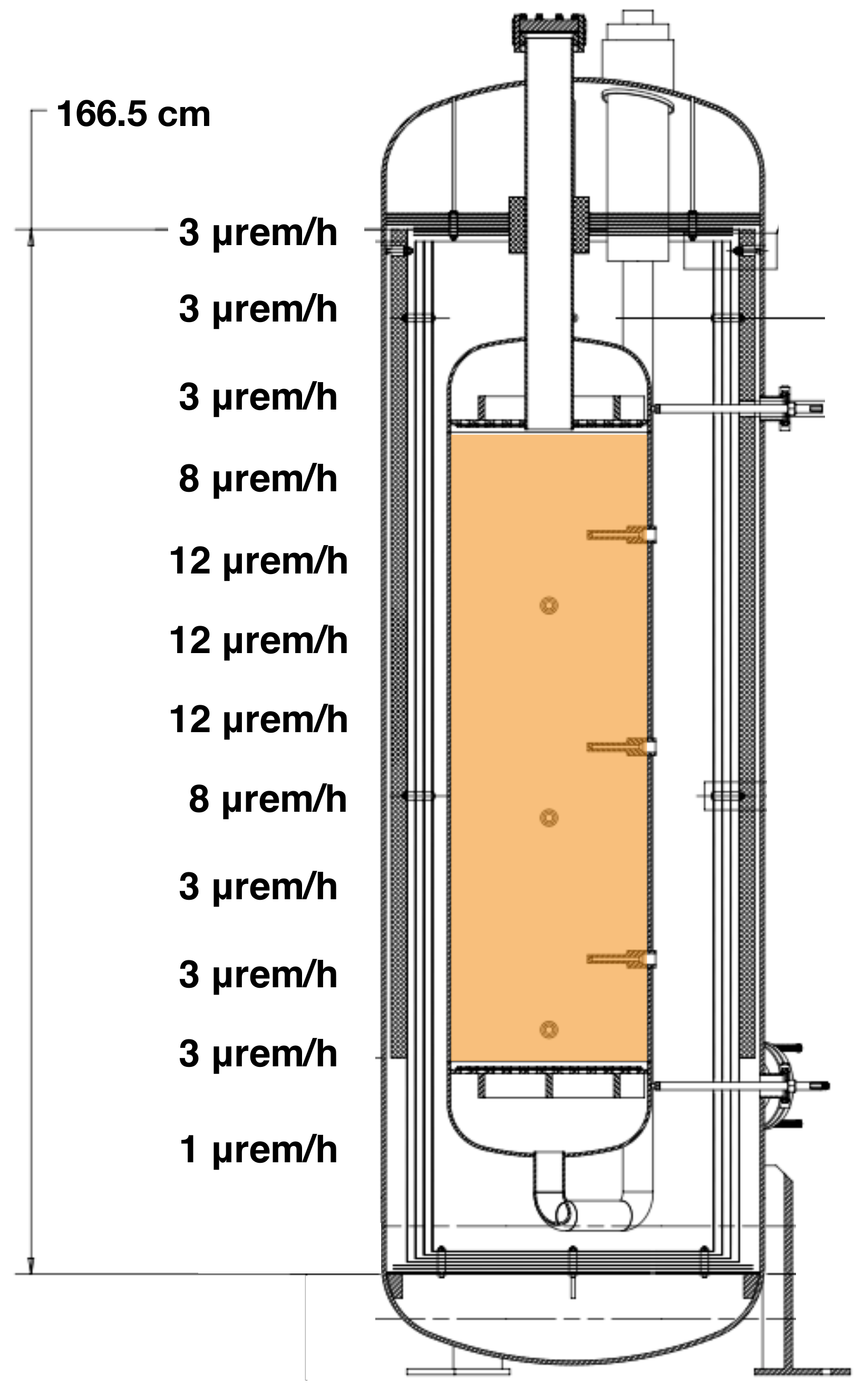}
    \caption{An engineering drawing of the copper filter used in MicroBooNE, with the active copper volume shaded. The radon activity readings were taken with a $\upmu$rem radiation detector taken at measured intervals down the filter. The readings are marked where they were taken and show stratification in the copper. The liquid argon is pumped from the top of the filter to the bottom.}
    \label{fig:survey}
\end{figure}

When the radon-doped LArTPC flows through the full-sized filter skid, we observed no appreciable increase in the rate of radon decays in the TPC. To determine where the radon was residing, we performed a radiological survey using a $\upmu$rem-sensitive survey meter. These data were collected during the full-sized skid data-taking period. The activity measured at the \ra~source housing was 15 $\upmu$rem/h, relative to background readings of 1--3 $\upmu$rem/h. In a complete
survey of the cryogenic system, background levels of radiation were observed in all components except the copper-based filter. Here, the activity measured was consistent with the source strength and showed a clear stratification within the 77~L copper filter, as shown in figure~\ref{fig:survey}. The mole sieve filter, located immediately upstream showed activity readings at background levels of radiation. After the LAr flow through these filters was stopped, no activity was observed in the mole sieve filter, while the same stratification shown in figure~\ref{fig:survey} was observed in the copper filter for several days. These findings suggest that the \rn~was able to pass through the mole sieve while the copper filter effectively removed the radon from the LAr flow.

After the radiological survey, we switched to using the $30\%$-sized filter skid since the distance at which radiation was measured in the 77~L copper filter exceeded the length of the 23.5~L of copper in the second filter system. Before switching to using this smaller filter skid we measured that both the copper and mole sieve filters were reading at background levels. When we switched to using the $30\%$-sized filter skid we saw no substantial increase in the amount of radon decays in the TPC, as shown in figure~\ref{fig:radondata}. Further, upon completing a subsequent radiological survey, we found that the 25.2~L mole sieve continued to read at background levels and the 23.5~L copper filter read at activity levels similar to the 77~L copper filter. 

\section{Radon Filtration}
\label{sec:efficacy}

It is unclear from our data the specific mechanism by which the copper filter is blocking the $^{222}$Rn. The most likely mechanism would be that the $^{222}$Rn is being ``slowed'' and travels through the copper filter material with a reduced velocity relative to the argon.  As the $^{222}$Rn travels through the filter it decays, and would eventually punch through the filter material with an activity suppressed by its transit time.

Alternatively, the $^{222}$Rn could be being ``captured'' by the copper filter material via an undetermined mechanism, e.g. freeze out. If this were happening we would expect the $^{222}$Rn to be either captured in the filter material or to flow through immediately. This would mean we would see an immediate turn on of radon entering the detector but with a source strength suppressed relative to our expectation from our 500~kBq source.  

While the slowing mechanism seems the most likely one to the authors, a few observations qualitatively do not appear to support that hypothesis. Any process that slows the flow of radon relative to argon would likely be driven by the surface area of the material and the mole sieve filter material has greater than a factor of 4.5 more surface area per gram of filter material, but we did not observe any radiation from $^{222}$Rn emanating from the mole sieve. Second, over the course of 8~months, the stratification observed in figure~\ref{fig:survey} remained unchanged. Finally, the 30\%-sized filter is significantly thinner than the stratification observed in the full-sized filter, but no punch was observed in that filter. This modest yet potentially conflicting evidence causes us to maintain consideration of both a slowing and a trapping mechanism in the following analysis.  

\subsection{Efficacy via Filter Slowing}

\begin{figure}
    \centering
    \includegraphics[width=0.95\linewidth]{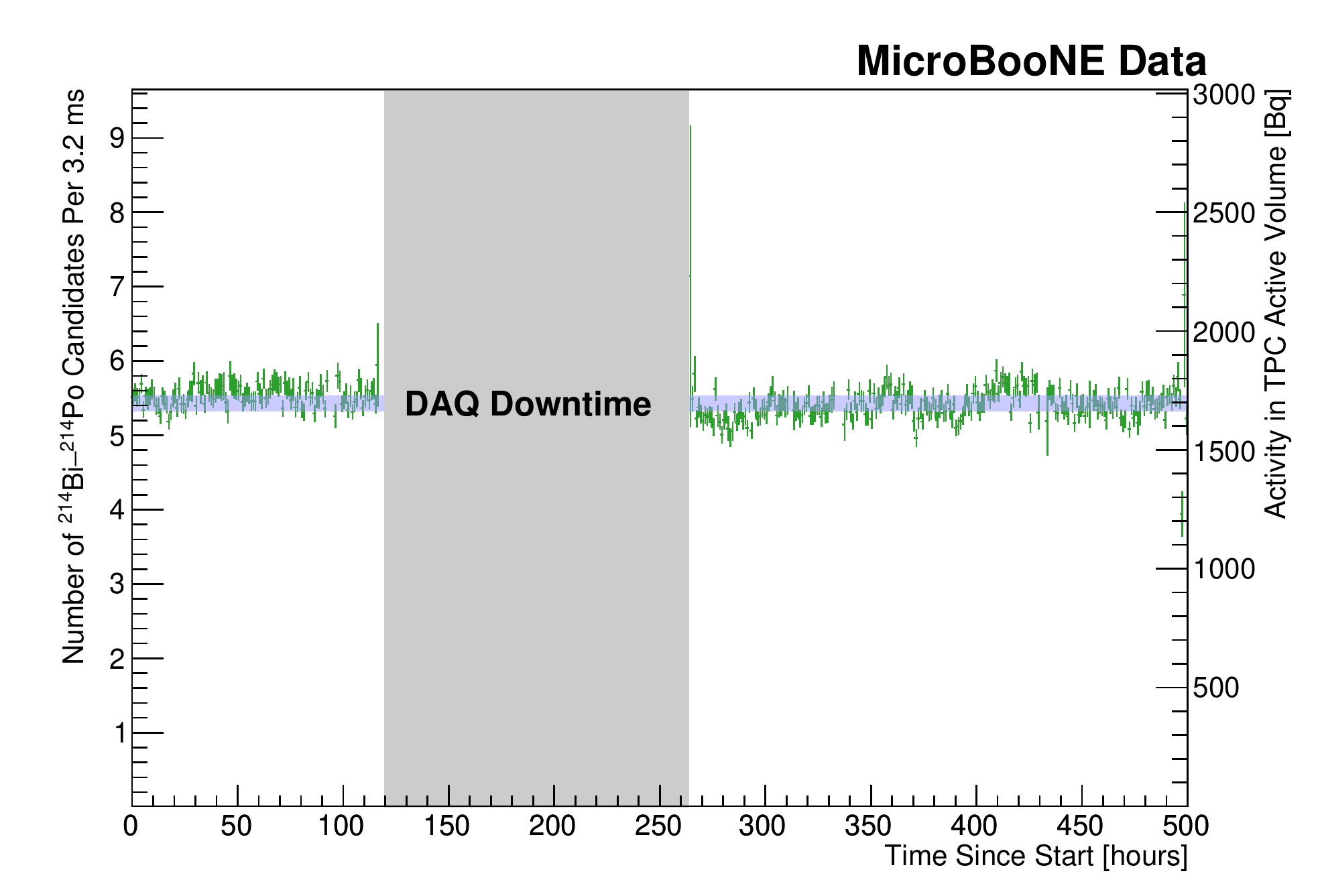}
    \caption{The average number of \bi--\po~coincident decays identified per readout (3.2~ms) versus time since radon was doped into the 30\%-sized filter. This activity rate can also be expressed in terms of Bq in the TPC active volume, shown on the right $y$-axis. The period marked with gray the DAQ was disabled but radon doping continued throughout this period. The blue band designates the fitted background level with its width including a 2\% systematic uncertainty on the background rate.}
    \label{fig:longrun}
\end{figure}

To assess the efficacy of the filter at slowing the $^{222}$Rn through the filter material one needs to take data for a long period of time. For this study, we left the $^{222}$Rn source flowing through the 30\%-sized filter skid for 500~hours. During this time there was a brief DAQ downtime where data was not recorded but the cryogenic and doping systems remained operable. Figure~\ref{fig:longrun} shows the number of \bi--\po~candidates observed during this period. Differently from the capture model, we are searching for a delayed shift in the baseline at any point after the beginning of doping. For this reason, we need to verify the stability of our background rates over this period of data taking. To do this we went back to data taken in 2018, without any radon source and in a similar detector state and compared the background rates. We found that the backgrounds from 2018 and now are consistent to the level of 2\% and use this as our systematic uncertainty on our background levels. 

We observed no significant shift in baseline over these 500~hours of data taking. At 500~hours the remaining radon that could be traveling through the filter will have decayed to $2.3\%$ of its original strength. This would be the equivalent of 11~kBq and we would observe a baseline shift up to $(9.4\pm2.9)$ events per 3.2~ms. Observing no shift in the baseline of this level we place a lower limit on the efficiency of our copper filter for ``slowing'' the radon of 97.7\%. Data taken over a longer period of time could help further constrain this value. 

\subsection{Efficacy via Filter Capture}

To determine the fraction of radon that is filtered by the MicroBooNE filtration system via capturing, a model of the expected activity as a function of time since the source was added, shown in figure~\ref{fig:rates}, is given by:
\begin{equation}\label{eq:rate}
\frac{dN}{dt} = B +
R^\mathrm{eff}_\mathrm{TPC} \cdot
\frac{\lambda_\mathrm{Rn222}}{\lambda_\mathrm{Rn222}-\lambda_\mathrm{Ra226}}
(e^{-\lambda_\mathrm{Ra226}t} - e^{-\lambda_\mathrm{Rn222}t})
\approx
B +
R^\mathrm{eff}_\mathrm{TPC} \cdot
(1 - e^{-\lambda_\mathrm{Rn222}t}),
\end{equation}

\noindent where $\lambda_\mathrm{Rn222}$ and $\lambda_\mathrm{Ra226}$ are the decay constants of $^{222}$Rn and $^{226}$Ra, respectively, $t$ is the time since the source was added, $B$ is the flat background term, and \refftpc~is the equilibrium activity in the TPC. Due to the size of $\lambda_\mathrm{Ra226} = 1.3\times10^{-11}$~s$^{-1}$ relative to $\lambda_\mathrm{Rn222}=1.2\times10^{-6}$~s$^{-1}$, we can safely ignore the former. Due to the shape of the MicroBooNE TPC relative to the cryostat, only 85~metric tons of LAr are within the TPC volume while the cryostat stores 175.2~metric tons of LAr, so the effective activity in the cryostat is $R^{\text{eff}}= 2.06 \times R^{\text{eff}}_{\text{TPC}} / \epsilon$, where $\epsilon$ is the selection efficiency (discussed in section~\ref{sec:sim}). The impact of the argon added to the cryostat as part of the doping has a negligible impact on the mass of LAr in the cryostat.   

Using the functional form of equation~\ref{eq:rate} we fit the average number of measured \bi--\po~coincident decays as a function of time ($dN/dt$ of equation~\ref{eq:rate}) across the full range of time shown in figure~\ref{fig:radondata} while floating the terms $B$ and $R^{\text{eff}}_{\text{TPC}}$. This procedure yields the effective source strength that we measure in the LArTPC active volume, which is then corrected by the selection efficiency and scaled from the LArTPC to the full cryostat, as discussed above. We marginalize over the background rate, which is treated as a nuisance parameter, and set a one-sided upper limit on $R^\text{eff}_\text{TPC}$. We then apply our efficiency correction, accounting for its uncertainty, to compute an upper limit on $R^\text{eff}$. The 2\% systematic fluctuations in the background rate over time, seen in figure~\ref{fig:longrun}, tend to pull the fit to the full time series to very low signal rates. Thus for the 30\% sized filter, we choose to conservatively fit only the first 120 hours, hence approximately half of the total source strength.
Fitting the data taken during the full-sized filter and the $30\%$-sized filter data results in limits of \reff~$<1.42$~Bq and $<0.61$~Bq, respectively.  
The resulting values of \reff~can then be compared to the 500~kBq source strength to estimate a lower bound on the fraction of radon that is removed by the filtration system, via $1-R^{\text{eff}}/500~\text{kBq}$. 
We find that the MicroBooNE filtration system, using the 77~L or 23.5~L copper filters, removes $>99.9997\%$ and $>99.9999\%$ of the radon introduced to the system, respectively.

\section{Implications}

These results have implications on the radon backgrounds and mitigation strategies for large scale LArTPCs. 
From the above analyses we find that our copper filter material successfully captures large fractions of the radon that is doped into the recirculated LAr. 
This would mean that placing a copper filter last in a filtering scheme could reduce any radon that enters the system upstream of this filter, similar to other radon traps~\cite{xent,charcoal}. 
An open question is whether or not the copper itself emanates any appreciable amount of radon, the analyses presented in this paper were insensitive to this. 
As a radon trap the copper filter could remove at least half the radon-based backgrounds, depending on where they originate from, if the filtration system can filter a full cryostat volume on a time-scale of 5.5~days. 
More R\&D into the radon mitigation mechanism, radon emanation, and filtration efficacy of copper filter material should be pursued to fully demonstrate the capabilities of copper as a radon mitigation system compatible with large liquid argon systems.  

\section{Conclusions}

We have tested the capability of the MicroBooNE filtration system to mitigate \rn~contamination, using a TPC-based topological reconstruction of \bi--\po~decay coincidences. Using reconstructed low-energy signatures in a single-phase LArTPC neutrino detector, we are able to identify the activity contributed by these radioactive decays. A 500~kBq \rn~source was used to test the efficacy of the MicroBooNE filter system to mitigate radon under two filtration hypotheses. If the radon were being slowed by the filter we found a lower limit of a 97.7\% reduction in the radon from the 23.5~L (20~kg) copper filter, with no observed radon appearing in the detector after 500~h. Alternatively, if the filter is capturing the radon we measured that $>99.9997\%$ and $>99.9999\%$ of the radon was removed by 77~L (65.5~kg) and 23.5~L (20~kg) copper filters, respectively.  A follow-up radiological survey revealed the presence of \rn~in the copper filter and the preceding mole sieve read at background radiation levels. This is the first observation of radon mitigation in liquid argon with a large-scale copper-based filter. These findings offer a potential solution for removing radon for future large LArTPCs.

\section*{Acknowledgements}

We gratefully acknowledge the discussions, feedback, and encouragement of Hugh Lippincott and Stephen Pordes throughout the course of the planning, execution, and analysis of our data.

This document was prepared by the MicroBooNE collaboration using the resources of the Fermi National Accelerator Laboratory (Fermilab), a U.S. Department of Energy, Office of Science, HEP User Facility. Fermilab is managed by Fermi Research Alliance, LLC (FRA), acting under Contract No.\ DE-AC02-07CH11359. MicroBooNE is supported by the following: the U.S. Department of Energy, Office of Science, Offices of High Energy Physics and Nuclear Physics; the U.S. National Science Foundation; the Swiss National Science Foundation; the Science and Technology Facilities Council (STFC), part of the United Kingdom Research and Innovation; the Royal Society (United Kingdom); and The European Union’s Horizon 2020 Marie Sk\l{}odowska-Curie Actions. Additional support for the laser calibration system and cosmic ray tagger was provided by the Albert Einstein Center for Fundamental Physics, Bern, Switzerland. We also acknowledge the contributions of technical and scientific staff to the design, construction, and operation of the MicroBooNE detector as well as the contributions of past collaborators to the development of MicroBooNE analyses, without whom this work would not have been possible.

\bibliographystyle{JHEP.bst}
\bibliography{bibliography.bib}

\end{document}